\documentstyle[12pt]{article}
\def\beq{\begin{equation}}
\def\eeq{\nonumber \end{equation}}

\def\mev{\,{\rm }}
\def\gev{\, {\rm GeV}}

\begin{document}

\title{The Inclusive Semileptonic $B$ Decay Lepton Spectrum from
$B \to X e \overline{\nu}$
\thanks{This work is supported
in part by funds provided by the U.S.
Department of Energy (DOE) under contract \#DE-AC02-76ER03069 and in
part by the Texas National Research Laboratory Commission
under grant \#RGFY92C6.\hfill\break
\vskip 0.05cm
\noindent $^{\dagger}$National
Science Foundation Young Investigator Award.\hfill\break
Alfred P.~Sloan
Foundation Research Fellowship.\hfill\break
Department of Energy Outstanding Junior
Investigator Award.\hfill\break
CTP\#2333, hep-ph/9407300\hfill June 1994}}
\renewcommand{\baselinestretch}{1.0}
\author{Lisa Randall$^{\dagger}$ \\
Massachusetts Institute of Technology\\
Cambridge, MA 02139\\
}
\date{}
\maketitle
\vskip-5in
December1993 \hfill  MIT-CTP\#2262
\vskip5in
\renewcommand{\baselinestretch}{1.2}
\abstract{
In this talk, we review the QCD calculations of the lepton spectrum from
inclusive semileptonic $B$ decay. We compare this prediction
to that of the ACCMM model. This latter work was
done in collaboration with Csaba Csaki.
\vskip 1in
{\sl Invited Talk, Presented at WHEPP-3 Workshop in Madras}
\thispagestyle{empty}
\newpage

In this talk we discuss the calculation of the inclusive semileptonic
decay lepton spectrum from $B \to X e \overline{\nu}$. We first
discuss the motivation for renewed interest in this calculation,
and discuss the use of the OPE and HQET to determine this spectrum.
We then compare this prediction to that of heavy quark models,
in particular that of Altarelli, Cabbibo, Corbo, Maiani, and Martinelli
\cite{accmm}
(hereafter referred to as ACCMM). I will concentrate on
$B \to X_c e \overline{\nu}$ and not the detailed questions
of the endpoint of the spectrum which are important for $B \to X_u e
\overline{\nu}$.

There are  several reasons for renewed interest in this prediction.
First of all is the copious production of $b$ quarks at LEP and CLEO.
A good understanding of the decay spectrum is necessary
for understanding fundamental $b$ quark parameters, which are
being precisely studied, such as the $b$ width and the forward--backward
asymmetry at LEP and those couplings relevant to the  rare
decays studied by CLEO. Furthermore, with a good understanding of the spectrum
one can do detailed measurements of heavy quark fragmentation,
which could give interesting tests of heavy quark theory and QCD
predictions \cite{nr,jr}. Furthermore, in order to do
high statistics measurements, one needs to do inclusive measurements.
On the theoretical side, it is of interest to study the
inclusive decays because they are under much better control
theoretically than predictions for exclusive modes, where
one needs to know hadronic matrix elements.
 An understanding of the lepton spectrum is therefore crucial to
extracting fundamental parameters.

In the seminal paper of Chay, Georgi, and Grinstein \cite{chay}
it was shown that
one  can treat the decay spectrum with the operator
product expansion  and heavy quark methods. They showed
the  leading $\Lambda/m$ corrections to the free quark result vanish
in the matrix elements. Subsequently,
 Bigi, Blok, Koryakh, Shifman, Uraltsev, and Vainshtein \cite{shif}
studied $(\Lambda/m)^2$ corrections. They
compared the  results to a free quark, ACCMM model
where they did not see the vanishing of the $\Lambda/m$ corrections
as manifest.
Following, there were further  calculations of $(\Lambda/m)^2$ suppressed
effects by
Falk, Luke, Savage  \cite{flukes} in the context of $b \to s\gamma$
and by Manohar and  Wise \cite{mw} in the context of semileptonic decays.
We will borrow heavily from these latter two papers in our review
of the QCD calculation.

\def\inv{^{\raise.15ex\hbox{${\scriptscriptstyle -}$}\kern-.05em 1}}
\def\pr#1{#1^\prime}  
\def\lbar{{\lower.35ex\hbox{$\mathchar'26$}\mkern-10mu\lambda}} 
\def\e#1{{\rm e}^{^{\textstyle#1}}}
\def\om#1#2{\omega^{#1}{}_{#2}}
%
%
%
%
\def\slash#1{\rlap{$#1$}/} 
\def\dsl{\,\raise.15ex\hbox{/}\mkern-13.5mu D} 
\def\delsl{\raise.15ex\hbox{/}\kern-.57em\partial}
\def\Ksl{\hbox{/\kern-.6000em\rm K}}
\def\Asl{\hbox{/\kern-.6500em \rm A}}
\def\Dsl{\hbox{/\kern-.6000em\rm D}} 
\def\Qsl{\hbox{/\kern-.6000em\rm Q}}
\def\gradsl{\hbox{/\kern-.6500em$\nabla$}}
\def\CAG{{\cal A/\cal G}}   
\def\CA{{\cal A}} \def\CB{{\cal B}} \def\CC{{\cal C}} \def\CD{{\cal D}}
\def\CE{{\cal E}} \def\CF{{\cal F}} \def\CG{{\cal G}} \def\CH{{\cal H}}
\def\CI{{\cal I}} \def\CJ{{\cal J}} \def\CK{{\cal K}} \def\CL{{\cal L}}
\def\CM{{\cal M}} \def\CN{{\cal N}} \def\CO{{\cal O}} \def\CP{{\cal P}}
\def\CQ{{\cal Q}} \def\CR{{\cal R}} \def\CS{{\cal S}} \def\CT{{\cal T}}
\def\CU{{\cal U}} \def\CV{{\cal V}} \def\CW{{\cal W}} \def\CX{{\cal X}}
\def\CY{{\cal Y}} \def\CZ{{\cal Z}}
%
%
%
%
%
\def\del{\partial}
\def\grad#1{\,\nabla\!_{{#1}}\,}
\def\gradgrad#1#2{\,\nabla\!_{{#1}}\nabla\!_{{#2}}\,}
\def\partder#1#2{{\partial #1\over\partial #2}}
\def\secder#1#2#3{{\partial^2 #1\over\partial #2 \partial #3}}
\def\ph{\varphi}
\def\bar#1{\overline{#1}}
\def\vev#1{\left\langle #1 \right\rangle}
\def\bra#1{\left\langle #1\right|}
\def\ket#1{\left| #1\right\rangle}
\def\abs#1{\left| #1\right|}
\def\vector#1{{\vec{#1}}}
\def\lform{\hbox{$\sqcup$}\llap{\hbox{$\sqcap$}}}
\def\darr#1{\raise1.5ex\hbox{$\leftrightarrow$}\mkern-16.5mu #1}
\def\dup{^{\vphantom{1}}}
\def\lie{\mathop{\hbox{\it\$}}} 
\def\half{{\textstyle{1\over2}}} 
\def\frac#1#2{{\textstyle{#1\over #2}}} 
%
%
%
%
\def\tr{\mathop{\rm tr}}
\def\Tr{\mathop{\rm Tr}}
\def\Im{\mathop{\rm Im}}
\def\Re{\mathop{\rm Re}}
\def\bR{\mathop{\bf R}}
\def\bC{\mathop{\bf C}}
\def\GeV{{\rm GeV}}
\def\MeV{{\rm MeV}}
\def\keV{{\rm keV}}
\def\eV{{\rm eV}}
\def\TeV{{\rm TeV}}
%
%
%
%

%
%
\def\ltap{\ \raise.3ex\hbox{$<$\kern-.75em\lower1ex\hbox{$\sim$}}\ }
\def\gtap{\ \raise.3ex\hbox{$>$\kern-.75em\lower1ex\hbox{$\sim$}}\ }
\def\gl{\ \raise.5ex\hbox{$>$}\kern-.8em\lower.5ex\hbox{$<$}\ }
\def\roughly#1{\raise.3ex\hbox{$#1$\kern-.75em\lower1ex\hbox{$\sim$}}}
%
%
\def\ie{\hbox{\it i.e.}}        \def\etc{\hbox{\it etc.}}
\def\eg{\hbox{\it e.g.}}        \def\cf{\hbox{\it cf.}}
\def\etal{\hbox{\it et al.}}
\def\dash{\hbox{---}}
\def\cok{\mathop{\rm cok}}
\def\coker{\mathop{\rm coker}}
\def\np#1#2#3{Nucl. Phys. B{#1} (#2) #3}
\def\pl#1#2#3{Phys. Lett. {#1}B (#2) #3}
\def\prl#1#2#3{Phys. Rev. Lett. {#1} (#2) #3}
\def\physrev#1#2#3{Phys. Rev. {#1} (#2) #3}
\def\ap#1#2#3{Ann. Phys. {#1} (#2) #3}
\def\prep#1#2#3{Phys. Rep. {#1} (#2) #3}
\def\rmp#1#2#3{Rev. Mod. Phys. {#1} (#2) #3}
\def\cmp#1#2#3{Comm. Math. Phys. {#1} (#2) #3}
\relax

\def\bbar{\overline {\rm b}}
\def\sbar{\overline {\rm s}}
\def\gone{\Gamma_1}
\def\gtwo{\Gamma_2}
\def\hbar{\bar h_Q}
\def\h{h_Q}
\def\qhat{\hat q}
\def\ms{m_s}
\def\mshat{\hat\ms}
\def\qhsl{\slash{\qhat}}
\def\vsl{\slash{v}}
\def\qsl{\hbox{/\kern-.5600em {$q$}}}
\def\ksl{\hbox{/\kern-.5600em {$k$}}}
\def\id{{\rm iD}}
\def\({\left(}
\def\){\right)}
\def\dtk{ {{\rm d}^3 k\over (2\pi)^3 2 E_\gamma} }
\def\dtp{ {{\rm d}^3 P_X\over (2\pi)^3 2 E_X} }
\def\qdotv{\qhat\cdot v}
\def\vdotq{v\cdot\qhat}
\def\lamone{\lambda_1}
\def\lamtwo{\lambda_2}
\def\bsg{{B\rightarrow X_s\gamma}}
\def\bsee{{B\rightarrow X_s\ell^+\ell^-}}
\def\gamleft{\frac12(1-\gamma_5)}
\def\gamright{\frac12(1+\gamma_5)}
\def\ssw{\sin^2\theta_W}
\def\km{V_{tb} V_{ts}^*}
\def\bmat#1{\langle B \vert {#1} \vert B \rangle}
\def\lqcd{\Lambda_{\rm QCD}}
\def\shat{\hat s}
\def\shatsq{\hat s^2}
\def\b{{\rm b}}
\def\s{{\rm s}}
\def\c{{\rm c}}
\def\u{{\rm u}}
\def\e{{\rm e}}
\def\O{{\cal O}}
\def\L{{\cal L}}
\def\d{{\rm d}}
\def\D{{\rm D}}
\def\im{{\rm i}}
\def\q{{\rm q}}
\def\vslash{v\hskip-0.5em /}
\def\Dslash{{\rm D}\hskip-0.7em /\hskip0.2em}
\def\mev{\,{\rm MeV}}
\def\gev{\,{\rm GeV}}
\def\ol{\overline}
\def\OMIT#1{}
\def\frac#1#2{{#1\over#2}}
\def\lamqcd{\Lambda_{\rm QCD}}
\def\etal{{\it et al.}}
\def\coeff#1#2{{\textstyle {#1\over#2}}}

Defining
$x={2 E_e \over m}$,
$\epsilon={m_f^2 \over m_b^2}$ and
$x_m= 1-\epsilon$, the free quark
decay spectrum is given by
\beq
{d \Gamma(m_b,E) \over d x}={G_F^2 m^5 \over 96\pi^3}{x^2 (x_m-x)^2 \over
(1-x)^3} \left[(1-x)(3-2x)+(1-x_m)(3-x)\right]
\eeq
The idea is then to justify and improve on this free quark result.
For the study of the inclusive decay, we sum over
all final states with given quantum numbers.
The spectrum and rate of the inclusive decay is  then
governed by short distance physics.
Intuitively,the justification is that in the $m \to \infty$ limit, the decay
time is much
shorter than hadronization time scale. With the OPE and HQET,
this can be made precise. The crucial observation is that
there is a  large range of mass scales in the final state
$m_D^2\le P_X^2\le m_B^2$ so that the
energy flowing through the hadron system scales with the mass
of the decaying heavy quark.
In the $m\to\infty$ limit, this is much larger than the QCD scale.
Away from $P_X^2\approx m_D^2$, the internal quark is far from mass shell.
In this case, we expect the  OPE to be useful; it
will prove valid except near the endpoint.

The general idea is
to relate the square of the matrix element of interest to the
imaginary part of the time ordered product of currents.
One can then perturbatively
compute the time ordered product with the operator product
expansion (in the region of phase space where it is valid).
However unlike a standard OPE it is useful to expand
in the heavy quark mass rather than $q^2$ in order to apply
HQET.
 The coefficients of  the heavy quark operators are then obtained
by evaluating the time ordered product between quark and gluon states.
The time ordered product is evaluated in the end by taking
matrix elements  of the operators appearing in the OPE between meson states.
One can use the HQET to learn about these matrix elements.

Semileptonic decay is determined from the weak hamiltonian density:
\beq
H_W = -  V_{jb}\ {4 G_F\over \sqrt 2}\ \bar q_j \gamma^\mu P_L b\ \bar e
\gamma_\mu P_L \nu_e = -  V_{jb}\ {4 G_F\over \sqrt 2}\ J^\mu_j J_{\ell \mu},
\label{C}
\eeq
The inclusive differential decay rate for $H_b \to X_{u,c}
e\overline{\nu_e}$
(here we will be restricting attention to meson decay) is
governed by
\beq
W^{\mu\nu}_j = \left(2\pi\right)^3\sum_X \delta^4\left(
p_{H_b}-q-p_X\right)\langle H_b (v,s)|J^{\mu\,\dagger}_j\ket{X}
\bra{X}J^\nu_j\ket{H_b(v,s)}
\label{D}
\eeq
where
\beq
p_{H_b} = M_{H_b}v^\mu=m_bv^\mu+k^\mu \label{F}
\eeq
The $W^{\mu\nu}$ can be expanded in terms of five form factors
\beq
W^{\mu\nu}=-g^{\mu\nu} W_1 + v^{\mu}v^{\nu} W_2 - i
\epsilon^{\mu\nu\alpha\beta} v_\alpha q_\beta W_3 + q^\mu q^\nu W_4 +
\left(q^\mu v^\nu + q^\nu v^\mu\right) W_5,
\label{E} \eeq

One then relates this to a time ordered
product via Im $T^{\mu\nu}=-\pi W^{\mu\nu}$ where
\begin{eqnarray}
T^{\mu\nu}&=&-i\int d^4 x\ e^{-i q \cdot x} \sum_s
\langle H_b (v,s)| T\left( J^{\mu\,\dagger}\left(x\right)
J^\nu\left(0\right)\right) \ket{H_b(v,s)}\nonumber\\
&=&-g^{\mu\nu} T_1 + v^{\mu}v^{\nu} T_2 - i
\epsilon^{\mu\nu\alpha\beta} v_\alpha q_\beta T_3 + q^\mu q^\nu T_4 +
\left(q^\mu v^\nu + q^\nu v^\mu\right) T_5.\nonumber
\label{H}
\end{eqnarray}

Let us consider the analytic structure.
There are cuts corresponding to the decay of interest,
to the process $e \overline{\nu} H \to X$, and
 to $e^+\nu_e H \to X$. The idea is then
to do a perturbative calculation away from the physical
cut, which is always possible away from the endpoint.

As an example,we find the leading order result. This is done
explicitly in \cite{mw}.
The matrix element of the time Fourier transformed, time ordered
product of currents
\beq
-i\int d^4 x  e^{-i q \cdot x} T(J^{\mu\dagger} J^\nu)
\label{M}
\eeq
is
\beq
{1\over (m_b v - q  + k )^2 - m_j^2 + i \epsilon}\
\bar u\ \gamma^\mu\, P_L\ \left(m_b\slash v - \slash q +
\slash k + m_j\right)
\ \gamma^\nu\, P_L\ u ,
\label{N}
\eeq
Using
\beq
\gamma^\mu \gamma^\alpha \gamma^\nu = g^{\mu\alpha} \gamma^\nu +
g^{\nu\alpha} \gamma^\mu - g^{\mu\nu} \gamma^\alpha + i
\epsilon^{\mu\nu\alpha\beta}\gamma_\beta \gamma_5.
\label{O}
\eeq
one finds the order $k^0$ term is
\begin{eqnarray}
{1\over\Delta_0} &&\bar u\, \Bigl\{\  \left(m_b v - q\right)^\mu
\gamma^\nu + \left(m_b v - q\right)^\nu \gamma^\mu -  \left(m_b \slash v
- \slash q\right) g^{\mu\nu}\nonumber\\
&&\qquad - i \epsilon^{\mu\nu\alpha\beta} \left(m_b v - q\right)_\alpha
\gamma_\beta\ \Bigr\} P_L\, u,
\label{P}
\end{eqnarray}
where
\beq
\Delta_0 = (m_b v - q  )^2 - m_j^2 + i\epsilon
\label{Q}
\eeq

By replacing the matrix element with the $b$ quark
operator which yields this amplitude when evaluated
between $b$ quark states, one obtains
\begin{eqnarray}
{1\over\Delta_0}& &\Bigl\{ \left(m_b v - q\right)^\mu g^{\nu\lambda}
+ \left(m_b
v - q\right)^\nu g^{\mu\lambda} -
\left(m_b  v -  q\right)^\lambda
g^{\mu\nu}\nonumber\\& &\qquad - i \epsilon^{\mu\nu\alpha\lambda} \left(m_b v
- q\right)_\alpha\Bigr\}\ \bar b\, \gamma_\lambda P_L\, b.\nonumber
\label{R}
\end{eqnarray}
Because it is a current, one has
\beq
\bra{H_b (v,s) }\bar b\, \gamma^\lambda \, b\ket{H_b (v,s)}=v^\lambda,
\label{S}
\eeq
since $b$ quark number is an exact symmetry.

So we can evaluate the leading order contribution {\it exactly}. By taking
the matrix element of the OPE between $H$ states, we get
\begin{eqnarray}
T_1^0&=&{1 \over 2 \Delta_0}(m_b-q \cdot v)\nonumber\\
T_2^0&=&{1 \over \Delta_0} m_b\nonumber\\
T_3^0&=&{1 \over 2 \Delta_0}\nonumber
\label{T}
\end{eqnarray}
So to get the amplitude $W^{\mu\nu}$ which determines the spectrum,
we need the imaginary part of $T^{\mu\nu}$. This is readily obtained from
\begin{eqnarray}
{1\over \Delta_0}&=&\delta((m_bv-q)^2-m^2)\\
{1\over \Delta_0^2}&=&-\delta'((m_bv -q)^2-m^2)\\
{1\over \Delta_0^3}&=&{1\over 2}\delta''((m_bv-q)^2-m^2)
\end{eqnarray}
(Here we only use the first one for the  leading order term)

The mass shell condition above when integrated over phase
space leads to just the free quark decay distribution!

However, with these methods, one can also consider higher order corrections.
 The important result will be that $\lqcd/ m$ corrections vanish
and only two matrix elements, one of which is known, are required
to obtain $(\lqcd/m)^2$ corrections

We first introduce heavy quark fields and the heavy quark effective theory.
Recall
that in the heavy quark limit, the heavy quark mass and spin decouple from the
soft degrees of freedom.
The heavy quark state looks like a static point source of charge.
The  $b$ quark is almost on mass shell
$p=m_b v+k$.

Define the  heavy quark field
\beq
b_v = {(1 + \slash v)\over 2} e^{im_b v\cdot x} b + \ldots,
\label{A}
\eeq
with leading order lagrangian
\beq
{\cal L} = \bar b_v\,( i v \cdot D)\, b_v + \ldots.
\label{B}
\eeq

Now expand in terms of HQET operators. We follow \cite{flukes} who
give the leading term in terms of the standard quark field,
but the mass suppressed operators in terms of heavy quark fields.

\beq
    T\{\O^{\dagger},\O\}\mathrel{\mathop=^{\rm OPE}}
    {1\over m_b}\left[ \O_0+{1\over2m_b}\O_1
    +{1\over4m_b^2}\O_2+\ldots\right]
\label{a}
\eeq
\begin{eqnarray}
    \O_0&=&\ol\b\,\Gamma\,\b\,,\nonumber \\
    \O_1&=&\ol b_v\,\Gamma\,\id_\mu b_v\,,\nonumber\\
    \O_2&=&\ol b_v\,\Gamma\,\id_\mu\id_\nu b_v\nonumber
\label{b}
\end{eqnarray}
As before, at leading order, we have
\beq
    \langle B|\,\ol\b\gamma^\mu\b\,|B\rangle=2P_B^\mu
\label{d}
\eeq
Now consider the first order mass suppressed terms.
\beq
 \langle M|\,\O_1\,|M\rangle=
    \langle M|\,\ol h\,\Gamma\,\id_\mu h\,|M\rangle=
    \langle M|\,\ol h\,\Gamma v_\mu v\cdot\id h\,|M\rangle
\label{f}
\eeq
This vanishes by the equation of motion!
Notice this critically depends on our phase choice in defining
heavy quark field.

An arbitrary counterterm
\beq
\delta m \overline{b_v}b_v
\eeq
was taken to vanish.
With this choice of quark mass, all linear corrections to the
free quark result vanish!

At higher order in $1/m$, one needs the two matrix elements
\begin{eqnarray}
K_b &\equiv & -  \bra{H_b(v,s)} \bar b_v {(i
D)^2\over 2m_b^2}\, b_v\ket{H_b(v,s)}\nonumber\\
G_b &\equiv&  Z_b\,\bra{H_b(v,s)}
\bar b_v\, { g G_{\alpha\beta}\sigma^{\alpha\beta}\over 4m_b^2}\, b_v
\ket{H_b(v,s)},
\label{V}
\end{eqnarray}

The second matrix element is known because it is related to the known
spin dependent mass splittings.
The first is not known, and is a parameter to be determined.

So, to summarize, the leading order  result  for the rate
and spectrum is the free quark result.
The first  order corrections vanish.
At second order, there is one unknown coefficient and one which
has been determined.
This analysis has assumed we are  far from endpoint, where the OPE is valid.
Had we applied this near the  endpoint, higher dimension operators
would not be suppressed and it would be necessary to
average the spectrum over a range of energies.

One can compare this result to the predictions of models. It
is
interesting to see how models reflect and differ from these
general QCD predictions.
Consider for example the ACCMM model. Here I review
work of Csaba Cs\'aki and myself \cite{cr}, but
other interesting references are that by Grant  Baillie  \cite{baillie}
and Ref.  \cite{bsuv}.
One finds that one can
 always {\it define} a quark mass so that $1/m$ corrections vanish.
However, at $1/m^2$, the model would differ from the QCD result.

In the ACCMM model, one models
 a $B$ meson decay as {\it disintegration}.
There is a spectator quark with mass $m_{sp}$ and momentum distribution
$\phi(|p|)$.
\beq
\phi(|p|)={4 \over \sqrt{\pi} p_f^3}\exp\left(-{|p|^2\over p_f^2}\right)
\eeq

The  $b$ quark momentum is determined by kinematic constraints.
The lepton spectrum is determined from the  decaying
off shell $b$ quark.
Define
\beq
E_W=M_B-\sqrt{p^2+m_{sp}^2}
\eeq

The invariant mass of the $b$ quark  will then be
\beq
W^2=m_B^2+m_{sp}^2-2m_B\sqrt{p^2+m_{sp}^2}
\eeq
Define

\begin{eqnarray*}
x&=&{2 E_e \over m}\\
\epsilon&=&{m_f^2 \over m_b^2}\\
x_m&=& 1-\epsilon
\end{eqnarray*}
Recall the formula for free quark decay.
\beq
{d \Gamma(m_b,E) \over d x}={G_F^2 m^5 \over 96\pi^3}{x^2 (x_m-x)^2 \over
(1-x)^3} \left[(1-x)(3-2x)+(1-x_m)(3-x)\right]
\eeq
Then the lepton spectrum in the ACCMM model is
\begin{eqnarray}
{d \Gamma_B \over dE}&=&\int_0^{p_{max}} dp p^2 \phi(|p|) \int {1\over \gamma}
{d^2 \Gamma(W,E') \over dE' d \cos\theta}
dE'  \times \nonumber\\
&& \ \ \ d\cos\theta \int{d \cos \theta_p \over 2} \delta(E-\gamma E'-\gamma
\beta E' \cos \theta_p)\nonumber
\end{eqnarray}
\beq
{d \Gamma_B \over dE}=\int dp p^2 \phi(|p|)
{1 \over 2 \beta \gamma ^2} \int {d^2 \Gamma \over dE' d\cos \theta} d \cos
\theta' {dE' \over E'}
\eeq
Using $\beta \gamma^2=p E_W/W(p)^2$ yields
\beq
{d \Gamma_B \over dE}=\int dp p^2 \phi(|p|){W^2 \over 2p E_W}
\int_{E_-}^{E_{max}} {dE' \over E'} {d \Gamma(W,E') \over dE'}
\eeq
where
\begin{eqnarray}
E^{\mp}&=&{E W \over E_W\pm p}\\
E_{max}&=&min\{E_+,{W \over 2}(1-x_m)\}\\
p_{max}&=&{m_B \over 2}-{m_f^2 \over m_B^2}
\end{eqnarray}
Now do a heavy meson mass expansion.
\beq
{d \Gamma_B \over dE}=\int_0^{p_{max}} \phi(|p|) p^2 {W^2 \over 2p E_W} {G_F^2
W^4 \over 48 \pi^3}
\int_{E_-}^{E_+}{dE' \over E'}\left({2E \over W}\right)^2 \left(3-{4E' \over
W}\right)\nonumber
\eeq

We can explicitly evaluate the $E'$ integral, to get
\beq\label{wfcn}
{d \Gamma_B \over dE}={G_F^2 E^2 \over 24 \pi^3}\int_0^{p_{max}}{dp \over
E_W}\left(
6 E_W W^2-8 E E_W^2-{8 \over 3} p^2 E \right) p^2 \phi(|p|)\nonumber
\eeq
Take $p_f$ small.
\beq
{d \Gamma_B \over dE}={d \Gamma_B^0 \over dE}+{G_F^2 E^2 M_B^2 \over 12 \pi^3
\sqrt{\pi}}\left(8{ E \over m_B}-12\right){p_f \over m_B}+P({p_F^2\over m_B})^2
{d \Gamma_q \over dE}
\eeq
It looks like there are nonvanishing  linear corrections,
but that is because we expanded in the meson mass $m_B$, rather than the quark
mass $m_b$.
To do a heavy quark expansion, one needs to
 {\it define} a quark mass. This can be done so that the linear
correction vanishes.
\beq
{d \Gamma_B \over dE}={G_F^2 E^2 \over 24 \pi^3}\int_0^{p_{max}}{dp \over
E_W}\left(
6 E_W W^2-8 E E_W^2-{8 \over 3} p^2 E \right) p^2 \phi(|p|)
\eeq
If we define
\beq
m_b=\langle W(p) \rangle
\eeq
Up to quadratic terms in $p$,we have
\beq
\langle W \rangle =\langle E_W \rangle=m_b=m_B-\langle p \rangle
\eeq
\beq
\langle f(W,E_W) \rangle = f(m_B,m_B)-\langle p \rangle f'(m_B,m_B) +
O(p^2)=f(m_b,m_b)+O(p^2)
\eeq

We see that it was possible to define a $b$ quark
mass so that linear corrections
vanished because the $p$ dependence was all through $E_W$.
So we can eliminate linear corrections through a proper choice of $m_b$.
It is easy  to extend this to nonzero spectator mass.

\beq
\langle E_W \rangle=m_B-\langle \sqrt{m_{sp}^2+p^2}\rangle
\eeq
Now consider the total rate and the full spectrum.
\beq
\int {d\Gamma_B \over dE}=\int dp p^2 \int {d \cos \theta_p \over 2} \int dE'
{d \Gamma(W,E') \over E'} \delta(E-\gamma E'-\gamma \beta E' \cos \theta_p)
\eeq
However, the spectra themselves are very different near the endpoint.
\begin{eqnarray*}
{d \Gamma_B \over dE}&=&\int _0^{p_1}dp p^2 \phi(|p|) {W \over 2p}
\int_{E_-}^{E_+}
{d \Gamma(W,E') \over E'}\nonumber \\
&+&\int_{p_1}^{p_{max}} dp p^2 \phi(|p|) {W \over 2p} \int_{E_-}^{E_1}
{d \Gamma(W,E') \over E'} \nonumber
\end{eqnarray*}
where
$E_1=W(1-\epsilon)/2$ ,
$p_1=m_B/2(1-{m_f^2 \over m_B^2}-E$ for $m_{sp}=0$.
This spectrum extends up to $E_{max}={m_B-m_{sp} \over 2} \left( 1-
{m_f^2 \over (m_B-m_{sp})^2}\right)$

We see the ACCMM spectrum and the free quark spectrum deviate
due to two effects near the endpoint. The first term is
 not integrated up to $p \approx \infty$ and the
second term integrates over a rapidly falling spectrum so it
is no longer a good approximation to replace $\langle f(W) \rangle$
by $f(\langle W \rangle)$. The spectrum
extends beyond naive quark mass endpoint.

Since we have a specific model, we can  also investigate the
question of how large an energy interval must
be averaged over to get good agreement between the model
and the free quark prediction. In ref. \cite{cr} we used
the averaging function
\beq
{d \Gamma \over dE}(E_0)=\int{1 \over \sqrt{\pi}\Delta E}
e^{-\left({(E-E_0)^2
 \over \Delta E^2}\right)}
{d \Gamma(E) \over dE}(E)
\eeq
We expect to require an average of approximately
 $2 p_f$ to get $(p_f/m)^2$ agreement. This was appproximately correct.
We also checked  that the best fit quark mass is
generally very close to the exact quark mass, so the
deviation of the model from the free quark prediction is
probably not very important for practical purposes.

 At higher order, the predictions of this model will not agree
with general QCD predictions.
This is because gauge invariance  is not incorporated.
There are corrections to  the free quark result due to the
fact that terms proportional to $\langle p_0^2 \rangle$
and $\langle \vec{p}^2 \rangle $
both yield $1/m^2$ corrections.
Recall
that in the heavy quark theory the operator $\langle D_0^2 \rangle $
contributions only at higher order in the heavy quark mass expansion
by the equation of motion.
 In the ACCMM model, both contribute at the same order.
Obviously, the ACCMM model also doesn't give you spin dependent
gluon operator.
We conclude that although the model is probably adequate in practice (as would
be free quark decay) it does not properly incorporate QCD dynamics.

We conclude that there has
been much  advancement in our understanding of the lepton decay spectrum.
It is unclear at what level this will be tested.
Most of the large deviations from the free quark
prediction are in the endpoint region, where
one must smear to get reliable predictions.
Furthermore, the higher order correction involves an unknown parameter and
requires very accurate measurements. However, from
a practical point of view, when using
the spectrum to extract $b$ quark couplings, we see that the free quark
decay models $b$ quark decay very well. It might be
better to fit the spectrum to a free quark spectrum that
that of a model, such as the ACCMM model.
The vanishing of  $\lqcd/m$ corrections is very important
to this conclusion.

Acknowledgements: I thank my collaborator C. Cs\'aki. I
also thank the Institute for Theoretical Physics in Santa
Barbara where this writeup was completed.

\end{document}